# Observation and study of the $J/\psi \rightarrow \mu^+\mu^-$ in the ATLAS experiment at LHC

E. Rossi on behalf of the ATLAS Collaboration
*Department of Physics - Università di Napoli "Federico II" Via Cinthia, 80126 Napoli, ITALY*

The high cross section for the J/ψ production and the wide transverse momentum spectrum accessible to ATLAS give to these events a central role in the calibration of the ATLAS detector. The data collected during the first months of the 2010 run at $\sqrt{s}$ = 7 TeV p-p collisions (integrated Luminosity up to 290 $nb^{-1}$) show a clear signal with two muons in the final state. The reconstructed invariant mass agrees with the PDG value and the peak width is compatible with Monte Carlo expectations. The doubly differential cross section is measured with respect to the transverse momentum and rapidity.

## 1. INTRODUCTION

Measuring the J/ψ production and properties in ATLAS [1] is a crucial step both for understanding the detector performance and for performing measurements of various B-physics channels. The J/ψ resonance provides an excellent testing ground for studies of muon trigger and identification efficiencies, as well as momentum scale and resolution of muons with transverse momentum below 20 GeV. In this paper the results of studies on the $J/\psi \rightarrow \mu^+\mu^-$resonance with the ATLAS detector using data up to an integrated luminosity of 290 $nb^{-1}$ collected in $\sqrt{s}$ = 7 TeV proton-proton collisions at the LHC between the end of March and July 2010 are presented. A measurement of the production cross section $d\sigma/dp_T dy$ of the J/ψ in bins of transverse momentum ($p_T$) and rapidity (y) is also shown.

## 2. THE ATLAS DETECTOR

The ATLAS detector covers almost the full solid angle around the collision point with layers of tracking detectors, calorimeters and muon chambers. For the measurements presented in this paper, the trigger system, the Inner Detector tracking devices (ID) and the Muon Spectrometer (MS) are of particular importance. The ATLAS ID (full coverage in φ and $|\eta| < 2.5$) consists of a silicon Pixel detector (Pixel), a silicon strip detector (SCT) and a Transition Radiation Tracker (TRT), all immersed in a 2 T axial magnetic field. The ATLAS MS designed to detect tracks over a large region of $|\eta| <2.7$, consists of a large toroidal magnet with an average magnetic field of 0.5 T in the barrel and four technologies: Monitored Drift Tube chambers (MDT) and Cathode Strip Chambers (CSC) used as precision chambers, Resistive Plate Chambers (RPC) and Thin Gap Chambers (TGC) used as trigger chambers. The ATLAS detector has a three-level trigger system: Level 1 (L1), Level 2 (L2), and the Event Filter (EF). For these measurements, the trigger relies on the Minimum Bias Trigger Scintillators and a dedicated EF muon trigger is required to confirm the candidate.

## 3. RECONSTRUCTION AND FIT OF J/Ψ MASS SIGNAL

To evaluate the J/ψ mass, the parameters of the ID tracks associated with the muons are used. Background contributions are expected to primarily come from heavy flavour decays, muons from pion and kaon in-flight decays, fakes from misidentification of hadrons as muons, and Drell-Yan production. To ensure that collision events are selected, events passing the trigger selection are required to have at least 3 tracks associated with the same reconstructed primary vertex.

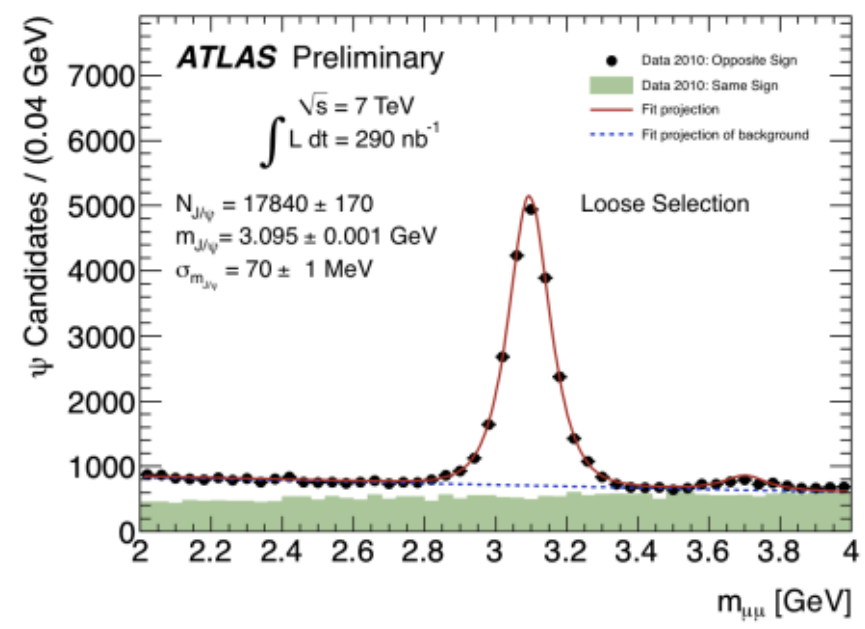

**Figure 1: Invariant mass distribution of reconstructed $J/\psi \rightarrow \mu+\mu-$ candidates after vertexing. Same sign combinations (green) are superimposed.**

The 3 tracks are required to have at least one hit in the pixel system and at least six hits in the SCT. In each surviving event, pairs of reconstructed muons are sought and only muons associated with ID tracks that have at least one hit in the pixels and six in the SCT are accepted. No $p_T$ cut is applied and at least one of the muons in each pair is required to be combined (i.e. the muon is reconstructed combining the track parameters of both a MS track and an ID track). The two inner detector tracks from each pair of muons passing these selections are fitted to a common vertex using the ATLAS offline vertexing tools based on the Kalman filtering method. No constraints on mass or pointing to the primary vertex are applied to the fit, and a very high vertex fit $\chi^2$ upper limit is applied ($\chi^2 < 200$). Those pairs which successfully form a vertex are regarded as $J/\psi \rightarrow \mu\mu$ candidates. Once the vertex fit is applied, the refitted track parameters and error matrices are used to calculate the invariant mass and per-candidate mass error. An unbinned maximum-likelihood fit is used to extract the $J/\psi$ mass and the number of $J/\psi$ signal candidates from the data. Figure 1 shows the invariant mass distribution of the opposite sign pairs after vertexing, with the same sign superimposed. The same sign pairs contribution in the J/ψ signal region is lower than the background from opposite sign pairs, as expected. The small excess in the background contribution from opposite sign pairs is in agreement with the expected contribution of muons from B and D meson decays in the current data. The kinematic properties of the event candidates in the J/ψ mass region have been studied (for details see [2] and [3]). MonteCarlo (MC) comparisons are made using two samples generated with PYTHIA 6, tuned using the ATLAS MC09 tune and MRST LO* parton distribution functions. For the MC J/ψ signal, it has been used an implementation of prompt J/ψ production subprocesses in the NRQCD Color Octet Mechanism framework, tuned to describe Tevatron results. For a description of the background the minimum-bias process, which includes all basic parton-parton scattering subprocesses has been used. To avoid any double-counting the prompt J/ψ production was removed from the minimum bias sample. The MC events were reconstructed with the same software used to process the data from the detector. The Invariant Mass fit mean value is 3.095 ± 0.001 GeV, in agreement with the PDG value for the J/ψ mass within statistical uncertainty. The signal resolution is 70 ± 1 MeV consistent with Monte Carlo.

## 4. DIFFERENTIAL INCLUSIVE J/Ψ CROSS SECTION

The comparison of the inclusive differential cross section of J/ψ production $d\sigma/dp_T dy$ measured by ATLAS, and the Monte Carlo simulations shows that the shapes of the distributions are largely in agreement in both $p_T$ and y, however the overall normalisation in the Monte Carlo is a factor of 10 higher (for details see [4]).

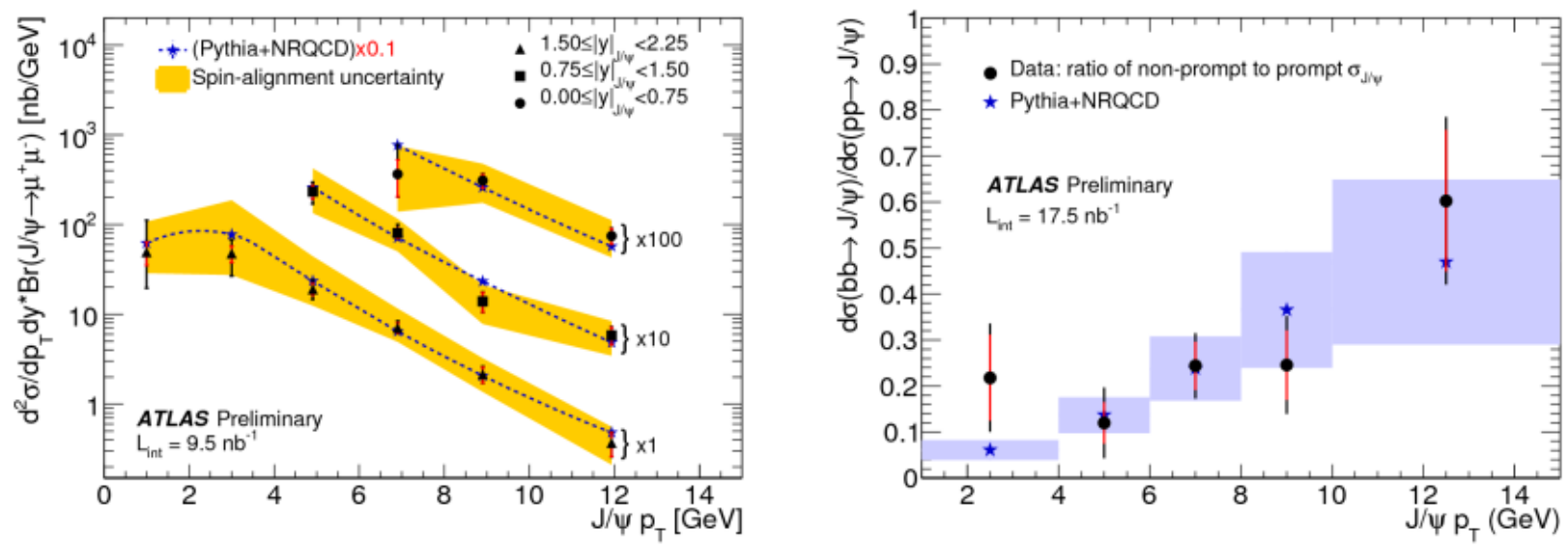

**Figure 2: Left plot shows the inclusive J/ψ production cross-section as a function of J/ψ $p_T$ and rapidity both for Monte Carlo and data. The yellow bands represent the variation of the results for the different spin-alignment scenarios considered. Right plot shows the ratio of non-prompt to prompt J/ψ production cross-sections as a function of J/ψ transverse momentum. The band represents PYTHIA MonteCarlo prediction.**

The results of the determination of the double differential cross section are shown in Figure 2 (left plot), with comparisons to PYTHIA predictions for ATLAS. Experimentally it is possible to distinguish between the J/ψs from decays of heavier charmonium states (prompt production) from the J/ψs produced in B hadron decays (non-prompt production). The former decays at their production point, which is the primary vertex of the event. The J/ψ mesons produced in B hadron decays will have a displaced decay vertex due to the long lifetime of their B hadron parent. From the measured distances between the primary vertices and corresponding J/ψ decay vertices we can infer the ratio of the production cross-sections of non-prompt J/ψ to promptly produced J/ψ. Shown in Figure 2 (right plot) are the results of the differential non-prompt to prompt ratio measurement, as a function of J/ψ $p_T$. The measured values of the ratio of production cross-sections are in good agreement with Monte Carlo expectations in the whole range of transverse momenta covered by this measurement, implying that the discrepancy with the overall normalisation from PYTHIA predictions is similarly affecting both prompt and non-prompt contributions.

## 5. SUMMARY AND CONCLUSIONS

The decay $J/\psi \rightarrow \mu^+\mu^-$ is observed in ATLAS data using combined information from the muon spectrometer and the inner detector. The Invariant Mass peak is in agreement with the PDG within the statistical uncertainty and the peak width is consistent with MC expectations. The double differential cross-section of inclusive J/ψ production has been measured, the behaviour shows agreement with Monte Carlo in both $p_T$ and y but the overall normalization factor in the Monte Carlo is a factor 10 . The ratio of non-prompt to prompt J/ψ has been measured as a function of the J/ψ transverse momentum between 1 –15 GeV. The measurement is in good agreement with Monte Carlo predictions.